\documentclass[10pt,twocolumn,a4paper]{article}

\usepackage[left=20mm, right=15mm, top=15mm, bottom=15mm]{geometry}

\usepackage{subfig}
\usepackage{flushend}
\usepackage{indentfirst}
\usepackage{graphics}
\usepackage{amsmath}
\usepackage{graphicx}
\usepackage{epstopdf}
\usepackage{float}

\usepackage[font=footnotesize,labelfont=bf]{caption}
\usepackage[labelsep=period]{caption}

\graphicspath{{./figure/}}

\begin{document}

\title{\huge \textbf{Calculations for electron-impact ionization of magnesium and calcium atoms in the method of interacting configurations in the complex number representation}}

\date{23.06.2017}

\twocolumn[
\begin{@twocolumnfalse}
\maketitle

\author{\textbf{V.M. Simulik}$^{1*}$, \textbf{Y-N.Y. Tsmur}$^{2}$, \textbf{R.V. Tymchyk}$^{1}$, \textbf{T.M. Zajac}$^{2}$\\\\

\footnotesize $^{1}$ {Institute of Electron Physics, National Academy of Sciences, Uzhgorod, 88000, Ukraine}\\

\footnotesize $^{2}${Department of Electronic Systems, Uzhgorod National University, Uzhgorod, 88000, Ukraine}\\

\footnotesize $^{*}$Corresponding Author: vsimulik@gmail.com}\\\\\\

\end{@twocolumnfalse}
]

\noindent \textbf{\large{Abstract}} \hspace{2pt} Next investigations in our program of transition from the He atom to the complex atoms description have been presented. The method of interacting configurations in the complex number representation is under consideration. The spectroscopic characteristics of the Mg and Ca atoms in the problem of the electron-impact ionization of these atoms are investigated. The energies and the widths of the lowest $^{1}S,\,^{1}P,\,^{1}D,$ and $^{1}F$ autoionizing states of Mg atom, and the lowest $^{1}P$ autoionizing states of Ca atom, are calculated. Few results in the photoionization problem on the $^{1}P$ autoionizing states above the n=2 threshold of helium-like Be$^{++}$ ion are presented.\\

\noindent \textbf{\large{Keywords}} \hspace{2pt} electron-impact ionization of atom, autoionizing states, quasistationary states, interacting configurations method\\

\noindent\hrulefill

\section{\Large{Introduction}}

We presented here next step in our last years program to apply the method of interacting configurations in the complex number representation (the ICCNR method) to the complex atoms description. The previous step was considered in [1, 2], where this method on the example of beryllium atom has been demonstrated.

The ICCNR method was suggested in papers [3-5] and successfully applied to the description of the quasistationary states of helium formed at its electron ionization in the energy interval above the threshold of excited ion formation. At the modern stage in the development of this method, a principal possibility is its application to the calculation of ionization processes in more complicated atomic structures.

Our step by step transition from the He atom description to the complex atoms consideration has been realized via the problem of ionization of H$^{-}$, Li$^{+}$ ions [6, 7] up to the enough complex atoms (such as Be, Mg and Ca) investigations. The results were reported at the international conferences [6, 8--11] as some approbation of the method and found data. The complete description of the method formalism was given in [12], see, e. g., [13] as well. In [14] the choice of the ground state wave function for such precision calculations of the quasistationary states parameters has been considered and discussed. One of the goals of these investigations is to demonstrate that the ICCNR method can be useful for the complex atoms study on the level of popular R-matrix approach, see, e. g., [15]. The comparison with theoretical calculations in other methods has been considered as well.

Thus, the ICCNR method is applied here to the calculation of spectroscopic characteristics of autoionizing states (AIS) of Mg and Ca atoms in the problem of the electron-impact ionization of these atoms. In particular, the energies and the widths of the lowest ($^{1}S,\,^{1}P,\,^{1}D,$ and $^{1}F$) AIS of Mg atom, and the lowest $^{1}P$ AIS of Ca atom, are calculated. Few results in the photoionization problem on the $^{1}P$ AIS above the n=2 threshold of helium-like Be$^{++}$ ion are presented as well. The first three $^{1}P$ resonances  above the n=2 threshold are considered. Found results are compared with [16, 17].

The exact quantum mechanical methods are welcome here because the experimental investigations of some atoms (e. g., beryllium atom) are complicated due to its chemical properties.

Some known results for Mg atom [18--20] are compared with our calculations in ICCNR method. Furthermore, our results (found in ICCNR method) for Ca atom are compered with experimental and theoretical investigations [21--23].

Note that the analysis of the loss spectrum of ejected electrons made it possible to compare indirectly the obtained results with the results of studies of the scattering problem.

As one can see in the literature, beryllium [16, 17, 19, 24--44], magnesium [18--20, 24, 25, 45--58] and calcium [21--23, 59, 60] atoms turns out to be the promising objects for researches.

\section{\Large{Some backgrounds of the method}}

The ICCNR method is a well-defined quantum-mechanical method for the calculation of parameters of atomic systems. This method is a development and a generalization of the known method of interacting configurations in the real number representation. It has some advantages in comparison with the standard method of interacting configurations in the real number representation and other calculation methods for the energies and widths of quasistationary atomic states. First, this is a capability of finding not only the energies, but also the widths of quasistationary states. Second, there are new possibilities for the resonance identification. The ICCNR method makes it possible, on the basis of the results of calculations, to estimate the contribution of each resonance state to the cross-section of the process and, if the resonance approximation is applicable, to introduce a set of parameters that determine the energies and the widths of quasistationary states, as well as the contours of resonance lines in the ionization cross-sections. In the concrete problems this approach also enables to investigate the applicability of approximate methods of cross-section estimation and to determine the limits of their validity. Those advantages make it possible to apply successfully the ICCNR method not only to scattering processes, but also to much more complicated processes such as ionization of atoms by electrons.

Consider the equation of the examined reaction
\begin{equation}
\label{eq1}
A(n_{0}L_{0}S_{0})+e^{-}(\overrightarrow{k}_{0}) \rightarrow A^{+}(nl_{1})+e^{-}(\overrightarrow{k}_{1})+e^{-}(\overrightarrow{k}),
\end{equation}

\noindent where $\overrightarrow{k}_{0},\overrightarrow{k}_{1},\overrightarrow{k}$ are the momenta of the incident, ejected, and scattered electrons, respectively. Then the generalized oscillator strength of the transition for the incident electron in the Born approximation is given by
\begin{equation}
\label{eq2}
\frac{df_{nl_{1}}}{dE}(Q)=\frac{E}{Q^{2}} \sum_{lL} |\langle nL_{1}El|\sum_{j=1}^{n}\exp(i\overrightarrow{Q}\overrightarrow{r_{i}})|n_{0}L_{0}S_{0}\rangle |^{2}.
\end{equation}

\noindent In this formula  $E=k_{0}^{2}-k^{2}$ is the energy loss, $\overrightarrow{Q}=\overrightarrow{k_{0}}- \overrightarrow{k}$ is the transmitted momentum, and $|nl_{1}El:LS_{0}\rangle$ is the wave function of an atom with total momentum $L$ and spin $S_{0}$ provided that an electron with momentum $l$ and energy $E$ is in the field of ion $A^{+}$, whose electron has the quantum numbers $|nl_{1}\rangle$. The function of the atomic ground state is given by $|n_{0}L_{0}S_{0}\rangle$.

Note that process (1) is a much more complicated physical phenomenon in comparison with the electron scattering by an atom. Exact theoretical calculations of such processes constitute a problem for modern theoretical physics. Therefore, the consideration of this problem for multielectron atoms in the framework of the ICCNR method is an important and challenging scientific step.

More details of the ICCNR method formalism can be found in [12].

\section{\Large{The results of calculations}}

Here the electron-impact ionizations of the Mg, Ca atoms in the interval of AIS excitation are considered. The corresponding results [1, 2] for beryllium atom are discussed briefly as well. Furthermore, some results for helium-like Be$^{++}$ ion are given.

\subsection{\normalsize \textbf{Energies and widths of helium-like Be$^{++}$ ion below the n=3 threshold autoionizing states of Be atom}}

In articles [1, 2] energies and widths of the lowest AIS ($^{1}S,\,^{1}P,\,^{1}D,$ and $^{1}F$) of a beryllium atom has been presented. These resonances were obtained in the ICCNR approximation in the problem of the electron-impact ionization of an atom. The indirect comparison with results of corresponding scattering problem has been fulfilled. Furthermore, the energies of $^{1}P$ states, which are located between the first and second ionization thresholds of a beryllium atom, are found and compared with the results of calculations obtained by other authors (see, e. g., [5--10] in [2]). In calculations, the Coulomb wave functions were used as basis configurations. For every term, up to 25 basis configurations were taken into account.

Here we are able to add energies and the widths in the photoionization problem of the $^{1}P$ AIS below the n=3 threshold of helium-like Be$^{++}$ ion. The first three $^{1}P$ resonances  above the n=2 threshold are presented. The results are compared with theoretical calculations of [16, 17].

\begin{table}[h]
\noindent\caption{Comparison of the energies
and the widths obtained with the use of the ICCNR
method for the AIS below the n=3 threshold of a helium-like Be$^{++}$ ion with the theoretical
results of other authors (the first three $^{1}P$ resonances 
above the n=2 threshold are under consideration)} 
\tabcolsep4.2pt
\begin{small}

\begin{tabular}{|c|c|c|c|c|c|}
\hline
\rule{0pt}{5mm} E, eV & $\Gamma$, eV& E, eV [16]& $\Gamma$, eV [16]& E, eV [17]& $\Gamma$, eV [17] \\
\hline
329.18 & 0.318   & 329.50 & 0.324  & 329.55 & 0.412 \\
\hline
333.24 & 0.0081   & 333.35 & 0.086  & 333.69 & 0.088 \\
\hline
337.47 & 0.0019   & - & -  & 337.66 & 0.0023 \\
\hline
\end{tabular}

\end{small}

\end{table}

\subsection{\normalsize \textbf{Electron-impact ionization of a Mg atom in the interval of the excitation
of autoionizing states}}

The investigation of the ionization of Mg atoms (and Mg$^{+}$ ions) by photons and electrons is a challenging problem. This assertion is proved by both experimental and theoretical papers of many authors (see, e.g., publications [24, 25, 45--58] cited here and papers [18--20], which are considered here directly). In brief publications [9, 10], we started to study the electron-impact ionization of a Mg atom in the AIS excitation interval with the use of the ICCNR method. In Table 2, the results of our calculations for the energies and the widths of the lowest AIS ($^{1}S,\,^{1}P,\,^{1}D,$ and $^{1}F$) of a Mg atom (obtained in the electron-impact ionization problem in the ICCNR approximation) are presented.

First, our results are compared with similar states that are formed in the problem of electron scattering
by Mg$^{+}$ ions [19] (see Table 2). Since another problem has been considered in article [19] -- namely,
the scattering one -- such a comparison is indirect. In paper [19] the calculations were carried out in the
diagonalization approximation. Second, in the framework of the problem of the electron-impact ionization
of atoms, the energies of $^{1}P$-states must coincide with those obtained in the problem of photoionization of
a Mg atom. Therefore, a direct comparison of our results with experimental ones [18] and with the results
of calculations on the basis of the 𝑅R-matrix method [20] can be made. In Table 3, the energy positions and
the widths calculated for the $^{1}P$𝑃 AIS of a magnesium atom with the use of the ICCNR method are directly compared with the experimental data of paper [18] and the theoretical data obtained with the help of the 𝑅R-matrix formalism [20], as well as with the problem of electron scattering by a Mg$^{+}$ ion [19].

Thus, the original scientific results obtained with the help of the ICCNR method [3--5] for the energies and the widths of the lowest AIS ($^{1}S,\,^{1}P,\,^{1}D,$ and $^{1}F$) of a Mg atom in the problem of electron-impact ionization of this atom are presented (see Table 2). Their novelty consists in the application of the exact calculation method, namely, the method of interacting configurations and, moreover, the ICCNR method. The comparison with the calculations of corresponding energies and widths of AIS carried out in the diagonalization approximation in the problem of electron scattering by Mg$^{+}$ ions (Table 2) is indirect (a different object in a different problem), but really testifies to the reliability of the results obtained. Some of the results obtained here, namely, the energy positions of the $^{1}P$𝑃 AIS of a Mg atom, can be directly compared with the experiment and the R-matrix calculations (see Table 3). The results of calculations carried out with the use of the ICCNR method are in good agreement with the corresponding calculations using the R-matrix method [20] and experimental results [18] (see Table 3).

\begin{table}[h]
\noindent\caption{Energies and widths of the lowest
AIS ($^{1}S,\,^{1}P,\,^{1}D,$ and $^{1}F$) of a Mg atom obtained
in the ICCNR approximation in the problem
of electron-impact ionization of an atom. In paper
[19], the energies of autoionizing states were
calculated in the diagonalization approximation
in the framework of the problem of electron
scattering by a Mg$^{+}$ ion}\tabcolsep4.2pt

\begin{center}

\begin{tabular}{|c|c|c|c|c|}
\hline
\rule{0pt}{5mm} $^{1}S$  & E, eV & $\Gamma$ eV &E, eV [19]& $\Gamma$ eV [19]\\
\hline
$4s^{2}$ & 13.08 & 0.0987  & 13.06 & 0.1010 \\
\hline
$3d^{2}$ & 14.61 & 0.0480  & 14.66 & 0.0502 \\
\hline
$4s5s$   & 14.92 & 0.0425  & 14.97 & 0.0473 \\
\hline
$4s6s$   & 15.48 & 0.0196  & 15.53 & 0.0185 \\
\hline
$3d4d$   & 15.59 & 0.0140  & 15.64 & 0.0129 \\
\hline
$4s7s$   & 15.78 & 0.0115  & 15.80 & 0.0107 \\
\hline
$4s8s$   & 15.80 & 0.0069  &   -   &    -   \\
\hline
$^{1}P$  & E, eV & $\Gamma$ eV   & E, eV [19]& $\Gamma$ eV [19] \\
\hline
$4s4p$   & 14.15 & 0.157   & 14.18 & 0.143  \\
\hline
$3d4p$   & 15.01 & 0.172   & 14.95 & 0.162  \\
\hline
$4s5p$   & 15.34 & 0.0324  & 15.29 & 0.0301 \\
\hline
$4s6p$   & 15.68 & 0.0682  & 15.64 & 0.0667 \\
\hline
$3d4f$   & 15.77 & 0.0481  & 15.74 & 0.0448 \\
\hline
$4s7p$   & 15.85 & 0.0059  & 15.86 & 0.0048 \\
\hline
$3s8p$   & 19.95 & 0.0140  & 19.93 & 0.0143 \\
\hline
$^{1}D$  & E, eV & $\Gamma$ eV  & E, eV [19]& $\Gamma$ eV [19] \\
\hline
$3d4s$   & 13.62 & 0.262   & 13.66 & 0.272  \\
\hline
$3d^{2}$ & 14.31 & 0.253   & 14.38 & 0.269  \\
\hline
$4d4s$   & 14.89 & 0.0192  & 14.96 & 0.0189 \\
\hline
$3d5s$   & 15.28 & 0.0869  & 15.30 & 0.0951 \\
\hline
$4p^{2}$ & 15.47 & 0.0570  & 15.49 & 0.0578 \\
\hline
$3d4d$   & 15.58 & 0.0865  & 15.55 & 0.0876 \\
\hline
$4s5d$   & 15.69 & 0.0258  & 15.66 & 0.0248 \\
\hline
$^{1}F$  & E, eV & $\Gamma$ eV   & E, eV [19]& $\Gamma$ eV [19] \\
\hline
$3d4p$   & 14.15 & 0.0225  & 14.66 & 0.0230 \\
\hline
$4s4f$   & 15.01 & 0.0110  & 15.28 & 0.0113 \\
\hline
$3d5p$   & 15.34 & 0.0540  & 15.53 & 0.0589 \\
\hline
$3d4f$   & 15.53 & 0.0052  & 15.63 & 0.0053 \\
\hline
$4s5f$   & 15.68 & 0.0201  & 15.71 & 0.0205 \\
\hline
$3d6p$   & 15.77 & 0.0104  & 15.88 & 0.0109 \\
\hline
$4s6f$   & 15.85 & 0.0125  & 15.90 & 0.0131 \\ [2mm]
\hline
\end{tabular}

\end{center}

\end{table}

\begin{table}[h]
\noindent\caption{Comparison of the energies
and the widths of the AIS of a magnesium atom
found with the use of the ICCNR method
with the experiment [18] and calculations
for $^{1}P$-states [20] (paper [20]: the photoionization
problem and the photoionization threshold;
article [19]: the scattering problem)}\tabcolsep4.2pt

\begin{tabular}{|c|c|c|c|c|}
\hline
\rule{0pt}{5mm} $^{1}P$  & E, eV  & $\Gamma$ eV & E, eV [19]& $\Gamma$ eV [19] \\
\hline
$4s4p$ & 14.15 & 0.157  & 14.18 & 0.143  \\
\hline
$3d4p$ & 15.01 & 0.172  & 14.95 & 0.162  \\
\hline
$4s5p$ & 15.34 & 0.0324 & 15.29 & 0.0301 \\
\hline
$3d5p$ & 15.53 & 0.0775 & 15.56 & 0.0758 \\
\hline
$4s6p$ & 15.68 & 0.00682& 15.64 & 0.00667\\
\hline
$3d4f$ & 15.77 & 0.0481 & 15.74 & 0.0448 \\
\hline
$4s7p$ & 15.85 & 0.00592& 15.86 & 0.00476\\
\hline
$4s8p$ & 15.90 & 0.0087 &   -   &    -   \\
\hline
$3d6p$ & 15.93 & 0.0295 &   -   &    -   \\
\hline
$4s9p$ & 15.95 & 0.0011 &   -   &    -   \\ [2mm]
\hline
\end{tabular}
\begin{tabular}{|c|c|c|c|c|c|}
\hline
\rule{0pt}{5mm} $^{1}P$  & E, eV &  $\Gamma$ eV & E, eV [20]& $\Gamma$ eV [20] & E, eV [18] \\
\hline
$4s4p$ & 14.15 & 0.157  & 14.2213 & 0.3921 & 14.18 \\
\hline
$3d4p$ & 15.01 & 0.172  & 14.9048 & 0.6078 &   -   \\
\hline
$4s5p$ & 15.34 & 0.0324 & 15.3133 & 0.0931 &   -   \\
\hline
$3d5p$ & 15.53 & 0.0775 & 15.7264 & 0.0890 & 15.24 \\
\hline
$4s6p$ & 15.68 & 0.00682& 15.6653 & 0.0142 & 15.61 \\
\hline
$3d4f$ & 15.77 & 0.0481 &    -    &    -   &   -   \\
\hline
$4s7p$ & 15.85 & 0.00592& 15.8675 & 0.0095 & 15.83 \\
\hline
$4s8p$ & 15.90 & 0.0087 & 15.9802 & 0.0111 & 15.98 \\
\hline
$3d6p$ & 15.93 & 0.0295 & 16.007  & 0.0417 &   -   \\
\hline
$4s9p$ & 15.95 & 0.0011 & 16.065  & 0.0019 & 16.06 \\ [2mm]
\hline
\end{tabular}

\end{table}

\subsection{\normalsize \textbf{Electron-impact ionization of a Ca atom in the interval of the excitation
of autoionizing states}}

The application of ICCNR method to calculate the lowest AIS of calcium atom was started in paper
[11]. The energies and the widths of the lowest $^{1}P$-states were calculated. The results were compared
with the data obtained by other authors. In Table 4, besides the results of our calculations, the experimental data [21] and the results of theoretical calculations [22, 23] are shown. Analysis of the presented data testifies that the classification of AIS proposed in work [22] is possible. Thus, the results of our calculations agree well with the theoretical data obtained by other authors.

\begin{table}[h]
\noindent\caption{Comparison of the energies and the widths obtained with the use of the ICCNR
method for the AIS of a Ca atom with the theoretical results of other authors and the experiment [21]}\tabcolsep4.2pt

\begin{center}

\begin{tabular}{|c|c|c|c|c|}
\hline
\rule{0pt}{5mm} $^{1}P$  & E, eV & E, eV [21]& E, eV [22]& E, eV [23] \\
\hline
$3d5p$ & 6.601 & 6.59   & 6.604 & 6.633 \\
\hline
$3d6p$ & 7.033 & 7.02   & 7.038 & 7.080 \\
\hline
$3d7p$ & 7.397 & 7.39   & 7.342 & 7.415 \\
\hline
$3d8p$ & 7.465 & 7.47   & 7.471 & 7.502 \\
\hline
$3d9p$ & 7.551 &   -    & 7.556 & 7.575 \\
\hline
$3d10p$& 7.610 &   -    & 7.614 & 7.624 \\
\hline
$4p5s$ & 7.159 & 7.13   & 7.166 & 7.300 \\
\hline
$3d4f$ & 6.937 &   -    & 6.938 & 6.960 \\
\hline
$3d5f$ & 7.240 & 7.25   & 7.248 & 7.260 \\
\hline
$3d6f$ & 7.425 &   -    & 7.427 & 7.427 \\
\hline
$3d7f$ & 7.523 &   -    & 7.529 & 7.527 \\
\hline
$3d8f$ & 7.591 &   -    & 7.596 & 7.593 \\
\hline
$^{1}P$& $\Gamma$ eV & $\Gamma$ eV [21]& $\Gamma$ eV [22]& $\Gamma$ eV [23] \\
\hline
$3d5p$ & 0.0801& 0.21   & 0.0702 & 0.0846  \\
\hline
$3d6p$ & 0.0059& 0.17   & 0.0056 & 0.0067  \\
\hline
$3d7p$ & 0.0451&   -    & 0.0509 & 0.0399  \\
\hline
$3d8p$ & 0.0261& 0.14   & 0.0232 & 0.0315  \\
\hline
$3d9p$ & 0.0163&   -    & 0.0141 & 0.0282  \\
\hline
$3d10p$& 0.0140&   -    & 0.0101 & 0.0207  \\
\hline
$4p5s$ & 0.0129& 0.15   & 0.0139 & 0.0132  \\
\hline
$3d4f$ &0.00006&   -    &0.000004& 0.00001 \\
\hline
$3d5f$ & 0.0059&   -    & 0.0028 & 0.00003 \\
\hline
$3d6f$ & 0.0019& 0.17   & 0.0014 & 0.0024  \\
\hline
$3d7f$ & 0.0009&   -    & 0.0011 & 0.00007 \\
\hline
$3d8f$ &0.00007&   -    & 0.00008& 0.00006 \\ [2mm]
\hline
\end{tabular}

\end{center}

\end{table}

\section{\Large{Conclusions}}

The method of interacting configurations in the complex number representation, which was applied earlier to the description of quasistationary states of a helium atom [3--5], is under consideration. The calculation of the ionization processes for more complicated atomic systems is suggested. The spectroscopic characteristics of the lowest AIS of Mg, Ca atoms were studied in the problem of the electron-impact ionization of these atoms (some results for helium-like Be$^{++}$ ion are presented as well). The energies and the widths of the lowest AIS ($^{1}S,\,^{1}P,\,^{1}D,\,^{1}F$) of Mg atom, and the lowest ($^{1}P$) AIS of Ca atom, were calculated. The found results were compared with known experimental data and calculations on the basis of other methods. Hence, we may draw conclusion about a successful verification of the ICCNR method for the calculation of AIS of complex atoms and the processes of electron-impact ionization and excitation of such atoms.


\begin{thebibliography}{99}

\small {

\bibitem{1} V.M. Simulik, T.M. Zajac, R.V. Tymchyk. Calculations for electron-impact ionization of beryllium in the method of interacting configurations in the complex number representations, arXiv: 1608.04078v1 [physics, atom-ph] 14 Aug. 2016. 5 p.

\bibitem{2} T.M. Zajac, V.M. Simulik, R.V. Tymchyk. The beryllium atom lowest autoionizing states in the method of interacting configurations in the complex number representations, Int. J. Theor. Math. Phys., Vol.6, No.4, 111--116 (2016).

\bibitem{3} S.M. Burkov, N.A. Letyaev, S.I. Strakhova, T.M. Zajac. Photon and electron ionization of helium to the n=2 state of He$^{+}$, J. Phys. B: At. Mol. Opt. Phys., Vol.21, No.7, 1195--1208 (1988).\\

\bibitem{4} S.M. Burkov, T.M. Zajac, S.I. Strakhova. Ionization of helium by fast electrons in the region over the threshold of exiting ions formation, Opt. Spektrosk., Vol.63, No.3, 17--25 (1988).\\

\bibitem{5} S.M. Burkov, S.I. Strakhova, T.M. Zajac. Total and partial generalized oscillator strengths for transitions to the continuum of helium, J. Phys. B: At. Mol. Opt. Phys., Vol.23, No.20, 3677--3690 (1990).\\

\bibitem{6} T.M. Zajac, A.I. Opachko, V.M. Simulik. Photoionization of Li$^{+}$ ion above the excited ion formation threshold. Book of Abstracts of 4-th CEPAS Conference, Cluj-Napoca, Romania, 18-20 June, 2008, p. 139.\\

\bibitem{7} T.M. Zajac, A.I. Opachko, V.M. Simulik. Application of the method of interacting configurations in complex number representation to the problem of the ionization of ions, Uzhgorod Univ. Sci. Herald, Ser. Phys., Vol.22, No.1, 82--86 (2008) (in Ukrainian).

\bibitem{8} A.I. Opachko, V.M. Simulik, R.V. Tymchyk, T.M. Zajac. Lowest autoionizing states of beryllium in photoionization problem. Book of abstracts of 43-th EGAS Conference, University of Fribourg, Fribourg, Switzerland, 28 June--2 July, 2011, p. 203.\\

\bibitem{9} T.M. Zajac, A.I. Opachko, V.M. Simulik, R.V. Tymchyk. Ionization of Mg atom by electron-impact in the region of autoionizing states excitation. Book of abstracts of 41-th EGAS Conference, University of Gdansk, Gdansk, Poland, 8--10 July, 2009, p. 156.\\

\bibitem{10} R.V. Tymchyk, T.M. Zajac, V.M. Simulik, A.I. Opachko. Calculation of energetic positions of the lowest autoionizing states of Mg atom. Abstracts of 10-th European Conference on Atoms Molecules and Photons, Salamanca, Spain, 4--10 July, 2010, Electron collisions, P-055.\\
	
\bibitem{11} V.M. Simulik, R.V. Tymchyk, T.M. Zajac. Autoionizing states of Ca in the problem of ionization of calcium atom by the electrons. Book of abstracts of 44-th EGAS Conference, University of Gothenburg, Gothenburg, Sweden, 9--13 July, 2012, p. 197.\\

\bibitem{12} T.M. Zajac, V.M. Simulik. The method of interacting configurations in complex number representation. Application for calculations of spectroscopic characteristics of quasi-stationary states in two electron systems, Int. J. Pure. Appl. Phys., Vol.3, No.3, 243--260 (2007).\\

\bibitem{13} V.M. Simulik, T.M. Zajac, R.V. Tymchyk. Application of the method of interacting configurations in the complex number representation to calculating the spectroscopic characteristics of the autoionizing states of Be, Mg and Ca atoms, Ukr. J. Phys., Vol.60, No.11, 1094--1100 (2015).

\bibitem{14} V.M. Simulik, T.M. Zajac, R.V. Tymchyk. Choice of the wave function for the helium ground state for precision calculations of quasistationary state parameters, Ukr. J. Phys., Vol.61, No.11, 950--955 (2016).

\bibitem{15} O. Zatsarinny, K. Bartschat, D.V. Fursa. Calculations for electron-impact excitation and ionization of beryllium, J. Phys. B: At. Mol. Opt. Phys., Vol.49, No.23, 235701(1--9) (2016).\\

\bibitem{16} Y.K. Ho. Autoionization states of helium isoelectric sequence below the n=3 hydrogenic thresholds, J. Phys. B: At. Mol. Opt. Phys., Vol.12, No.3, 387--400 (1979).\\

\bibitem{17} A. Wague. Application of the diagonalization approximation to the n=3 resonant photoionization of helium-like systems, Z. Phys D. At. Mol. Clust., Vol.6, No.4, 337--344 (1987).

\bibitem{18} M.A. Baig, J.P. Connerade. Extensions to the Spectrum of Doubly Excited MgI in the Vacuum Ultraviolet, Proc. Roy. Soc. Lond. A., Vol.364, December 20, 353--366 (1978).\\

\bibitem{19} V.L. Lengyel, V.T. Navrotsky, E.P. Sabad. Resonant scattering of low-energy electrons by Be$^{+}$ and Mg$^{+}$ ions, J. Phys. B: At. Mol. Opt. Phys., Vol.23, No.16, 2847--2867 (1990).\\

\bibitem{20} D.S. Kim, S.S. Tayal. Autoionizing resonances in the photoionization of ground state atomic magnesium, J. Phys. B: Atom. Mol. Opt. Phys., Vol.33, No.17, 3235--3247 (2000).\\

\bibitem{21} J.E. Kontrosh, I.V. Chernyshova, O.B. Shpenik.  Near-threshold electron-impact ionization of calcium atom, Opt.Spectrosc., Vol.110, No.4, 500--507 (2011).\\

\bibitem{22} O.I. Zatsarinny, V.I. Lengyel, E.A. Masalovich. Resonance structure in the electron-impact excitation of Ca$^{+}$ below the 5\textit{s} threshold, Phys. Rev. A., Vol.44, No.11, 7343--7354 (1991).\\

\bibitem{23} P. Scott, A.E. Kingston, A. Hibbert. Photoionization of neutral calcium, J. Phys. B: Atom. Mol. Phys., Vol.16, No.21, 3945--3959 (1983).\\

\bibitem{24} G. Mehlman-Ballofet, J.M. Esteva. Far ultraviolet absorption spectra with auto-ionized levels of beryllium and magnesium, Astrophys. J., Vol.157, No.2, 945--956 (1969).\\

\bibitem{25} J.M. Esteva, G. Mehlman-Ballofet, J. Romand. Spectres d'absorption dans l'ultraviolet lointain de Be, B, C, N, Mg, Al et Si, Journal of Quantitative Spectroscopy and Radiative Transfer, Vol.12, No.9, 1291--1303 (1972).\\

\bibitem{26} P.L. Altic. Photo-ionization cross section of beryllium near threshold, Phys. Rev., Vol.169, No.1, 21--26 (1968).\\

\bibitem{27} H.C. Chi, K.N. Huang, K.T. Cheng. Autoionizing levels of beryllium from the multiconfiguration relativistic random-phase approximation, Phys. Rev. A., Vol.43, No.5, 2542--2545 (1991).\\

\bibitem{28} D.S. Kim, S.S. Tayal, H.L. Zhou, S.T. Manson. Photoionization of atomic beryllium from the ground state, Phys. Rev. A., Vol.61, No.6, 062701--062709 (2000).\\

\bibitem{29} T.M. Clark et al. Observation of autoionizing states of beryllium by resonance-ionization mass spectrometry, J. Opt. Soc. Am. B., Vol.2, No.6, 891--896 (1985).\\

\bibitem{30} R Moccia, P Spizzo. One-photon transition probabilities and photoionisation cross section calculations of Be, J. Phys. B: At. Mol. Phys., Vol.18, No.17, 3537--3554 (1985).\\

\bibitem{31} H. Bachau, P. Galan, and F. Martin. Feshbach-model potential approach for the study of resonant
and bound states of Be-like ions, Phys. Rev. A., Vol.41, No.7, 3534--3544 (1990).\\

\bibitem{32} N. Miura, Y. Osanai, T. Noro, F. Sasaki. Theoretical determination of energies and widths of
autoionizing states of the Be atom, J. Phys. B: At. Mol. Opt. Phys., Vol.29, No.13, 2689–2699 (1996).\\

\bibitem{33} K. Bartschat, P.G. Burke, M.P. Scott. Electron-impact excitation of beryllium, J. Phys. B: At. Mol. Opt. Phys., Vol.29, No.20, L769--L772 (1996).\\

\bibitem{34} J. Colgan et al. Electron-impact ionization of all ionization stages of beryllium, Phys. Rev. A., Vol.68, No.3, 032712(1--9) (2003).\\

\bibitem{35} R. Wehlitz, D. Lukic, J. B. Bluett. Resonance parameters of autoionizing Be 2\textit{pnl} states, Phys. Rev. A., Vol.68, No.5, 052708(1--5) (2003).\\

\bibitem{36} C.P. Ballance et al. Electron-impact excitation of beryllium and its ions, Phys. Rev. A., Vol.68, No.6, 062705(1--11) (2003).\\

\bibitem{37} M.A. Uddin et al. Electron-impact ionization of beryllium isoelectronic ions, Int. J. Mass. Spectr., Vol.244, No.1, 76--83 (2005).\\

\bibitem{38} J.C. Cardona, J.L. Sanz-Vicario. Autoionizing states in beryllium-like atomic systems using explicitly correlated coordinates, J. Phys. B: At. Mol. Opt. Phys., Vol.41, No.5, 055003(1--10) (2008).\\

\bibitem{39} M.R. Talukder. Electron impact total ionization cross sections of beryllium and boron isoelectronic ions, Appl. Phys. B., Vol.93, No.2-3, 567--574 (2008).\\

\bibitem{40} W.C. Chu, C.D. Lin. Theory of ultrafast autoionization dynamics of Fano resonances, Phys. Rev. A., Vol.82, No.5, 053415(1--9) (2010).\\

\bibitem{41} M.S. Pindzola, C.P. Ballance, F. Robicheaux, J. Colgan. Electron-impact double ionization of beryllium, J. Phys. B: At. Mol. Opt. Phys., Vol.43, No.10, 105204(1--5) (2010).\\

\bibitem{42} J.M.N. Djiokap, A.F. Starace. Resonant enhancement of the harmonic generation spectrum of beryllium, Phys. Rev. A., Vol.88, No.5, 053412(1--15) (2013).\\

\bibitem{43} T. Maihom et al. Electron-impact ionization cross sections of beryllium and beryllium hydrides, Eur. Phys. J. D., Vol.67, No.1, 2(1--5) (2013).\\

\bibitem{44} O.I. Zatsarinny, K. Bartschat, D.V. Fursa, I. Bray. Calculations for electron-impact excitation and ionization of beryllium, arXiv: 1606.01326v1. [physics. atom-ph] 4 Jun 2016, 9 p.\\

\bibitem{45} Y.L. Shao, C. Fotakis, D. Charalambidis. Multiphoton ionization of Mg in the wavelength region of 300-214 nm, Phys. Rev. A., Vol.48, No.5, 3636--3643 (1993).\\

\bibitem{46} M.J. Ford et al. Electron-impact double ionization of magnesium, Phys. Rev. A., Vol.57, No.1, 325--330 (1998).\\

\bibitem{47} K. Bartschat, D. Weflen, X. Guan. Electron-impact ionization of magnesium, J. Phys. B: Atom. Mol. Opt. Phys., Vol.40, No.16, 3231--3239 (2007).\\

\bibitem{48} T.K. Fang, Y.K. Ho. Determination of resonance energies and widths of Mg 3\textit{pnl} $^{1}$D$^{e}$ and $^{1}$F$^{0}$ doubly excited states by the stabilization method with the B-spline-based configuration interaction approach, J. Phys. B: Atom. Mol. Opt. Phys., Vol.32, No.15, 3863--3872 (1999).\\

\bibitem{49} A. Reber, F. Martin, H. Bachau, R.S. Berry. Two-photon above-threshold ionization of magnesium, Phys. Rev. A., Vol.65, No.6, 063413(1--7) (2002).\\

\bibitem{50} A. Reber, F. Martin, H. Bachau, R.S. Berry. Three-photon above-threshold ionization of magnesium, Phys. Rev. A., Vol.68, No.6, 063401(1--10) (2003).\\

\bibitem{51} A. Reber et al. Above-threshold ionization near the 3\textit{p}4\textit{d} $^{1}$\textit{F}$^{0}$ autoionizing state in magnesium, Phys. Rev. A., Vol.71, No.5, 053402(1--7) (2005).\\

\bibitem{52} D. Rassi, V. Pejcev, T.W. Ottley, K.J. Ross. High-resolution ejected-electron spectrum of magnesium autoionising levels following two-electron excitation by low-energy electron impact, J. Phys. B: Atom. Mol. Phys., Vol.10, No.14, 2913--2921 (1977).\\

\bibitem{53} T.M. El-Sherbini, A.A. Rahman. Autoionizing states in MgI, Annalen der Phys., Vol.39, No.5, 333--337 (1982).\\

\bibitem{54} T.N. Chang. 3\textit{pnp} $^{1}$\textit{S} autoionization states of the magnesium atom, Phys. Rev. A., Vol.36, No.11, 5468--5470 (1987).\\

\bibitem{55} C.J. Dai. Resonance profiles of Mg 3\textit{pns} autoionizing states, Chin. Phys. Lett., Vol.12, No.3, 152--155 (1995).\\

\bibitem{56} R.F. Boivin, S.K. Srivastava. Electron-impact ionization of Mg, J. Phys. B: Atom. Mol. Opt. Phys., Vol.31, No.10, 2381--2394 (1998).\\

\bibitem{57} N.J. Kylstra, H.W. van der Hart, P.G. Burke, C.J. Joachain. Singly, doubly and triply resonant multiphoton processes involving autoionizing states in magnesium, J. Phys. B: Atom. Mol. Opt. Phys., Vol.31, No.14, 3089--3116 (1998).\\

\bibitem{58} J.A. Ludlow et al. Electron-impact single ionization of Mg and Al$^{+}$, Phys. Rev. A., Vol.79, No.3, 032715(1--6) (2009).\\

\bibitem{59} S.M. Kazakov, O.V. Khristoforov. Electron spectra from autoionizing states of strontium and calcium excited by low and intermediate-energy electrons, Sov. Phys. JETP, Vol.61, No.4, 656--664 (1985).\\

\bibitem{60} D. Cvejanovic, A.J. Murray. Single ionization of calcium by electron impact, J. Phys. B: Atom. Mol. Opt. Phys., Vol.36, No.17, 3591--3605 (2003).\\

}

\end{thebibliography}
\end{document}